\documentclass{book}    % fundamental class file is 'book'.
\usepackage{piers}  % use the file 'piers.sty'.
\begin{document}

%=== Input Title here ================================
\title{Beam-Shaping PEC Mirror  Phase   Corrector Design      } \maketitle
%===================================================

%=== List of authors (in order) ========
%-- Author(s) for the first affiliation ---
\author      {F. M. Lastname}
\affiliation {University}
\address     {}% optional
\city        {Boston}
\postalcode  {}% optional
\country     {USA}
\phone       {345566}    % optional
\fax         {233445}    % optional
\email       {email@email.com}  % optional
\misc        { }  % optional
\nomakeauthor
%------------------------------------

%=== List of authors (in order) ========
%-- Author(s) for the second affiliation ---
\author      {F. M. Lastname}
\affiliation {University}
\address     {}% optional
\city        {Boston}
\postalcode  {}% optional
\country     {USA}
\phone       {345566}    % optional
\fax         {233445}    % optional
\email       {email@email.com}  % optional
\misc        { }  % optional
\nomakeauthor
%-------------------------------------
%=== List of other authors (in order) ========
%......

%---Output of Authors----------------------
\begin{authors}

{\bf Shaolin Liao and Ronald J. Vernon}    \\
%\medskip
Department of Electrical and Computer Engineering  \\ 1415
Engineering Drive, Univ. of Wisconsin, Madison, U.S.A., 53706
\end{authors}
%--------------------------
%---Content of Paper Abstract-----------------------
\begin{paper}

\begin{piersabstract}
 The Perfect Electric Conductor (PEC) mirror phase corrector plays an important role in the beam-shaping
   mirror system design for   Quasi-Optical (QO) mode converter (launcher) in
    the sub-THz
    high-power
  gyrotron. In this article, both the Geometry Optical (GO) method and the
  phase gradient method have been presented for   the PEC mirror phase corrector design.
   The advantages and   disadvantages are discussed for both methods.
   An efficient algorithm has been proposed for  the phase gradient method.
\end{piersabstract}

  \begin{center}
  {\bf I. Introduction}
\end{center}

 The PEC mirror phase corrector is essential to shape the input beam from  the QO mode
converter (launcher) into the desired Fundamental Gaussian Beam
(FGB) in the sub-THz high-power gyrotron
\cite{Rong,Perkins,Shaolin_30,Shaolin_31_1,Shaolin_31_2,Shaolin_JEMWA,  liao_near-field_2006, shaolin_liao_new_2005,  liao_beam-shaping_2007, liao_fast_2007, liao_validity_2007, liao_high-efficiency_2008, liao_four-frequency_2009, vernon_high-power_2015, liao_multi-frequency_2008, liao_fast_2007-1, liao_miter_2009, liao_fast_2009, liao_efficient_2011, liao_spectral-domain_2019}.
  Figure \ref{fig:setup} shows the diagram of such beam-shaping
  mirror   system, in which the 4 pieces of PEC mirrors (M$_1$, M$_2$, M$_3$ and M$_4$)
  serve as the  phase correctors, aiming at shaping the
  input beam from the QO launcher into the desired FGB output beam. During  the iterative
   beam-shaping mirror system design   \cite{Rong,Shaolin_JEMWA}, phase unwrapping   is commonly required
  in the PEC mirror phase corrector design, which can effectively suppress  the
  edge diffraction   due to the discontinuities if  otherwise  the
  wrapped phase is used. In this article, both the GO method and the phase gradient
  method are discussed. An FFT-based efficient
 algorithm is also proposed to   speed up the PEC mirror phase corrector design  for the phase gradient
 method. The time dependence $e^{i \omega t}$ is assumed.

\begin{figure}[h] \centering
\includegraphics[width= 3.6 in, height= 2.7  in]{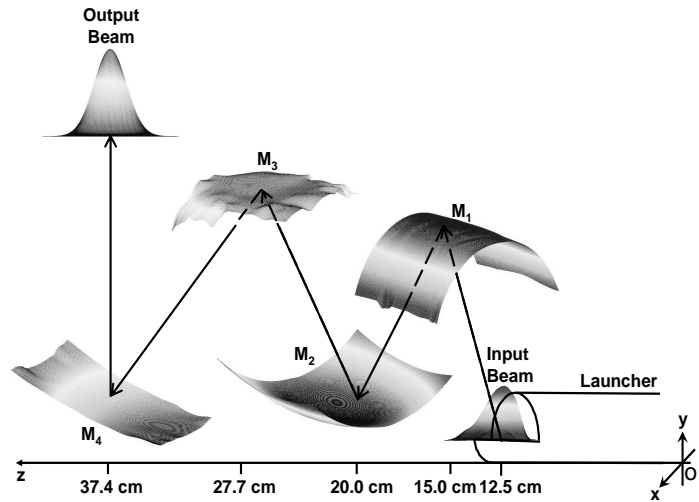}
\caption{ The diagram of the beam-shaping mirror system consisting
of 4 pieces of PEC mirrors to shape the input beam from the QO
launcher into the desired FGB output beam. The approximate {\bf z}
coordinates of the 4 pieces of PEC mirrors have been marked on {\bf
z} axis.}
 \label{fig:setup}
  \end{figure}

  \begin{center}
  {\bf II. The Problem of Phase Unwrapping}
\end{center}

 The phase correction requires the knowledge  of the unwrapped  2-Dimensional (2D) phases of the incident electric
 field  and the reflected electric   field  ($ \theta^i$, $\theta^r$). However,
 the phase obtained from the electric
 field {\bf E} through $\check{\theta} = \arctan\left[\frac{\Im({\bf E})}{\Re({\bf E})}\right]$
 ($\Re$ and $\Im$ denote the real part and the
imaginary part respectively) is the  wrapped  2D
  phase, which contains discontinuities of $2 n \pi$ (n is an integer). So, in order to ensure
the smoothness  of the PEC mirror surface, the wrapped 2D phases
($\check{\theta}^i$, $\check{\theta}^r$) must be unwrapped through
the 2D phase unwrapping
 methods \cite{Gerchberg,Ghiglia}.

  Mathematically, in the ideal situation where there is no residues in the wrapped  2D  phase
  $\check{\theta}$, the discreet phase gradient
   $\nabla  {\theta} = \nabla \check{\theta}$ (assuming that $\nabla  {\theta} < \pi$) and the 2D phase
 unwrapping can be expressed as,

\vspace{-0.2in}

{
\begin{eqnarray}\label{path_integral}
 \theta  = \int_C \nabla \check{\theta}   \cdot d{\bf r} + \theta({\bf
r}_0)
\end{eqnarray}
}

 \hspace{-0.23in}where, $ \theta$ denotes the 2D unwrapped phase along the integration path $C$ and $r_0$ denotes the starting point of the
 path  integration. Note that  the unwrapped phase $\theta$  obtained through (\ref{path_integral}) should not
  depend on the integration path $C$.   However, due to the residues in practice, the discrete phase gradient
  should be written as $\nabla \check{\theta} = (\nabla
g + \nabla \times {\bf R})$ and the unwrapped phase $\theta$ is
obtained as follows,

{
\begin{eqnarray}\label{solenoid}
 \theta  = \int_C (\nabla g + \nabla \times {\bf R})
\cdot d{\bf r} + \theta({\bf r}_0)
\end{eqnarray}
}

 From (\ref{solenoid}), it can be seen that $\nabla \times {\bf R} \ne
 0$ is caused by the existence of residues  and the unwrapped phase  $\theta$ depends on the
 integration path $C$.
 There are many 2D phase unwrapping
algorithms to deal with the residues in the literatures
\cite{Gerchberg}, \cite{Ghiglia}.  For example, the
  path following algorithm (e.g., ``quality-guided" method and ``mask-cut" method) gives
  faithful congruent unwrapped phase (with $2 n \pi$ difference from the wrapped phase).
However,  path following algorithm is time-consuming and the
unwrapped phase contains many discontinuities due to the existence
of residues. Another commonly-used algorithm, the minimum norm
method unwraps the wrapped phase by minimizing the $r$-norm phase
difference
  between the gradients of the wrapped phase and the desired unwrapped phase
\cite{Ghiglia},

\vspace{-0.1in}

{
\begin{eqnarray}\label{norm}
 Q  = \sum_X          \sum_Z    \left[  \hspace{-0.15in} \begin{array}{cccc}  \\  \\ \\  \end{array} w_x  \left| \frac{\partial \theta}{\partial
 x} -  \frac{\partial \check{\theta}}{\partial x}
  \right|^{r} + w_z \left| \frac{\partial \theta}{\partial z}  -
  \frac{\partial \check{\theta}}{\partial z}  \right|^{r}
   \hspace{-0.15in} \begin{array}{cccc}  \\  \\ \\  \end{array}
   \right]
  \end{eqnarray}

}

\hspace{-0.23in}where, $w_x$ and $w_z$ are weights for $\hat{\bf x}$
and $\hat{\bf z}$ directions respectively. When $r=2$, it is called
the Least Mean  Square (LMS) method.

\begin{figure}[b] \centering
\includegraphics[width= 3.5  in, height= 2.6 in]{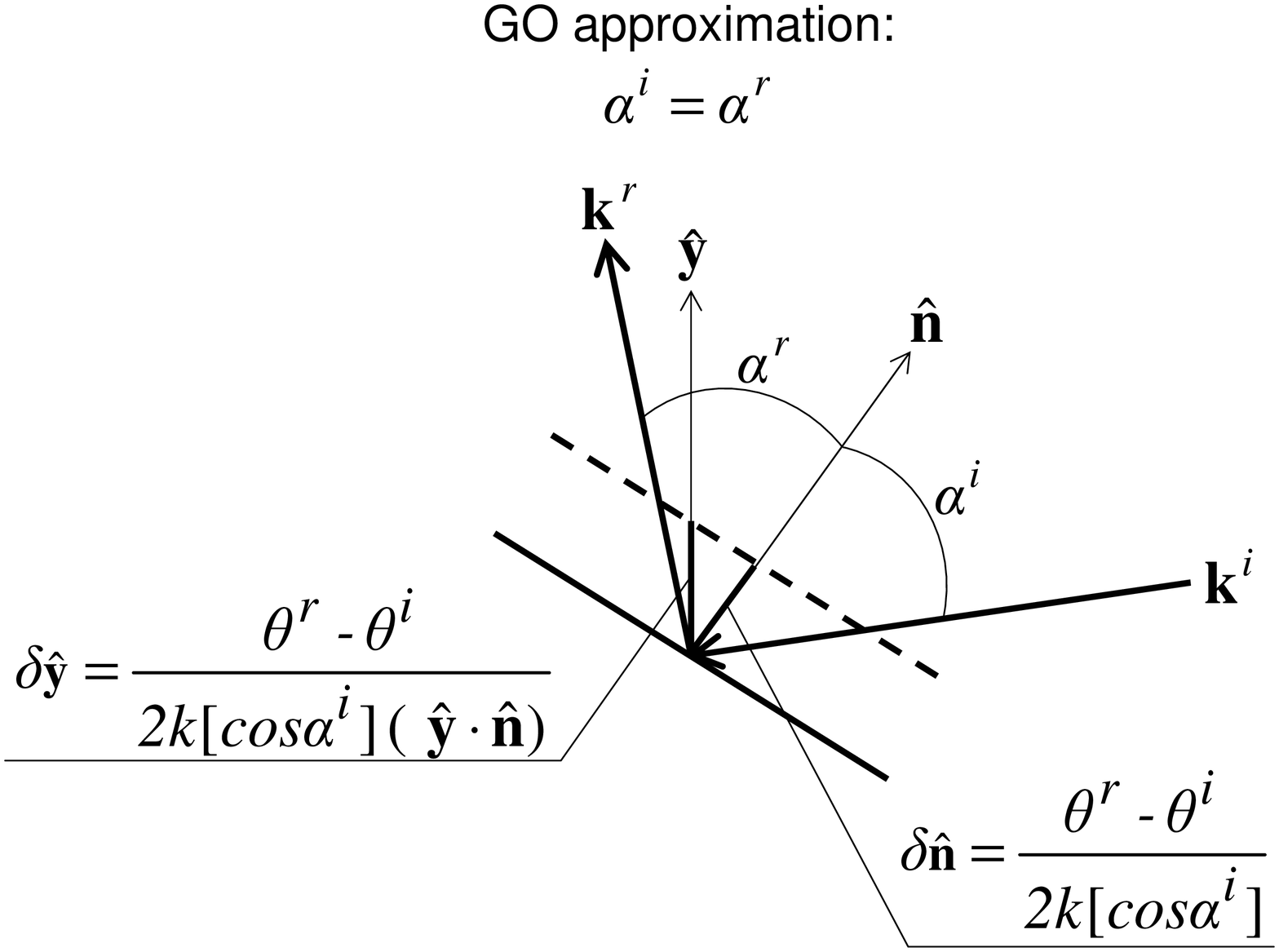}
\caption{ The PEC mirror surface correction in the sub-THz QO
regime: ${\bf k}^i$ and ${\bf k}^r$ are  wave vectors for the local
incident beam (with incident angle  $\alpha^i$) and the local
reflection beam (with reflected angle $\alpha^r$).   $\delta_{
\hat{\bf n}}$ is the PEC mirror surface correction in $\hat{\bf n}$
direction  and $\delta_{\hat{\bf y}}$ is the PEC mirror surface
correction in $\hat{\bf y}$ direction. }
 \label{surface_correction}
\end{figure}

  \begin{center}
  {\bf III. The GO Method}
\end{center}

In the sub-THz QO regime, it is reasonable to assume that the
  intensity or magnitude of the electric field is locally constant and
 the  local phase change can be evaluated through the GO method, as shown in Fig. \ref{surface_correction}.
 For fixed computational grid given on x-z plane (in favor of FFT
 operation), $\delta_{\hat{\bf y}}$ is preferred, which is rewritten
 as follows ($\cos \alpha^i =   \frac{   {\bf k}^i \cdot
\hat{\bf n} }{k}$),

 \begin{eqnarray}\label{deltay}
\delta_{\hat{\bf y}} = \frac{\delta \theta}{ 2   \left( {\bf k}^i
\cdot \hat{\bf n} \right) \left( \hat{\bf y} \cdot \hat{\bf n}
\right)}, \ \ \ \ \ \ \ \  \delta \theta = \theta^r - \theta^i
\end{eqnarray}

There are two approaches to calculate the local   wave vector ${\bf
k}$ (incident wave vector ${\bf k}^i$ and reflected wave vector
${\bf k}^r$), i.e., 1) the Poynting vector approach; and 2) the
phase gradient approach. The Poynting vector approach assumes that
the local beam propagates in the direction given by the Poynting
vector,

\vspace{-0.1in}

\begin{eqnarray}\label{poyting}
 {\bf k}  \propto {\bf E} \times \left({\bf H}  \right)^{\ast}  \propto  {\bf E}  \times
 \nabla \times \left({\bf E}  \right)^{\ast}
\end{eqnarray}

The phase gradient approach approximates the local wave vector as
the gradient of the phase,

\begin{eqnarray}\label{gradient}
 {\bf k}  \propto  \nabla \theta
\end{eqnarray}

It is not difficult to show that the two approaches are equivalent
in the far-field limit.

  \begin{center}
  {\bf IV. The Phase Gradient Method }
\end{center}

Instead of (\ref{deltay}), the expression of the PEC mirror surface
correction $\delta_{\hat{\bf y}}$ in the phase gradient method is
given as

\vspace{-0.1in}

  \begin{eqnarray}\label{phase_gradient}
\delta_{\hat{\bf y}} = \frac{\delta \theta}{ \nabla \left( \delta
\theta \right) } = \frac{\delta \theta}{ \nabla \theta^r - \nabla
\theta^i }
\end{eqnarray}

The phase gradient $\nabla \theta$ for the electric field ${\bf E} =
|{\bf E}| e^{i \theta}$ can be found as

  \begin{eqnarray}\label{gradient1}
   \nabla {\bf E} & = & \nabla \left\{ |{\bf E}|  e^{i \theta}  \right\}
    \\
  & = & \nabla \left\{ |{\bf E}|  \right\}   e^{i \theta} +   |{\bf E}|
    \nabla \left\{    e^{i \theta} \right\} \nonumber  \\
    & = &  \left[ \hspace{-0.12in} \begin{array}{cc} \\  \\ \end{array}  \nabla \left\{ |{\bf E}|  \right\}   +  i |{\bf
    E}|
    \nabla   \theta   \hspace{-0.12in} \begin{array}{cc} \\  \\ \end{array}  \right]  e^{i \theta}    \nonumber
\end{eqnarray}

  \begin{eqnarray}\label{gradient2}
   \rightarrow  \nabla \theta  & = & \Im\left[ \frac{\nabla {\bf E} }{{\bf E}  }
   \right] = \Im\left[ \hspace{-0.12in} \begin{array}{cc} \\  \\ \end{array} \nabla \ln{\bf
   E} \hspace{-0.12in} \begin{array}{cc} \\  \\ \end{array}
   \right]  \\
 \rightarrow  \nabla \left\{ |{\bf E}|  \right\}   & = & \Re\left[  \nabla {\bf E}   e^{ - i \theta}
   \right]   \nonumber
\end{eqnarray}

 From (\ref{gradient1}) and (\ref{gradient2}), the expression for
$\nabla \left( \delta \theta \right)$ in (\ref{phase_gradient}) is
obtained,

  \begin{eqnarray}\label{delta_phase}
  \nabla \left( \delta \theta \right)  & = &  \Im\left[  \frac{\nabla {\bf E}^r }{{\bf E}^r  } - \frac{\nabla {\bf E}^i }{{\bf E}^i
  }
   \right] = \Im\left[ \hspace{-0.12in} \begin{array}{cc} \\  \\ \end{array}
   \nabla \ln \frac{ {\bf    E}^r}{ {\bf E}^i} \hspace{-0.12in} \begin{array}{cc} \\  \\ \end{array}
   \right]
\end{eqnarray}

 \begin{figure}[h]
 \centering
  \includegraphics[width=6.9  cm, height=8.4cm ]{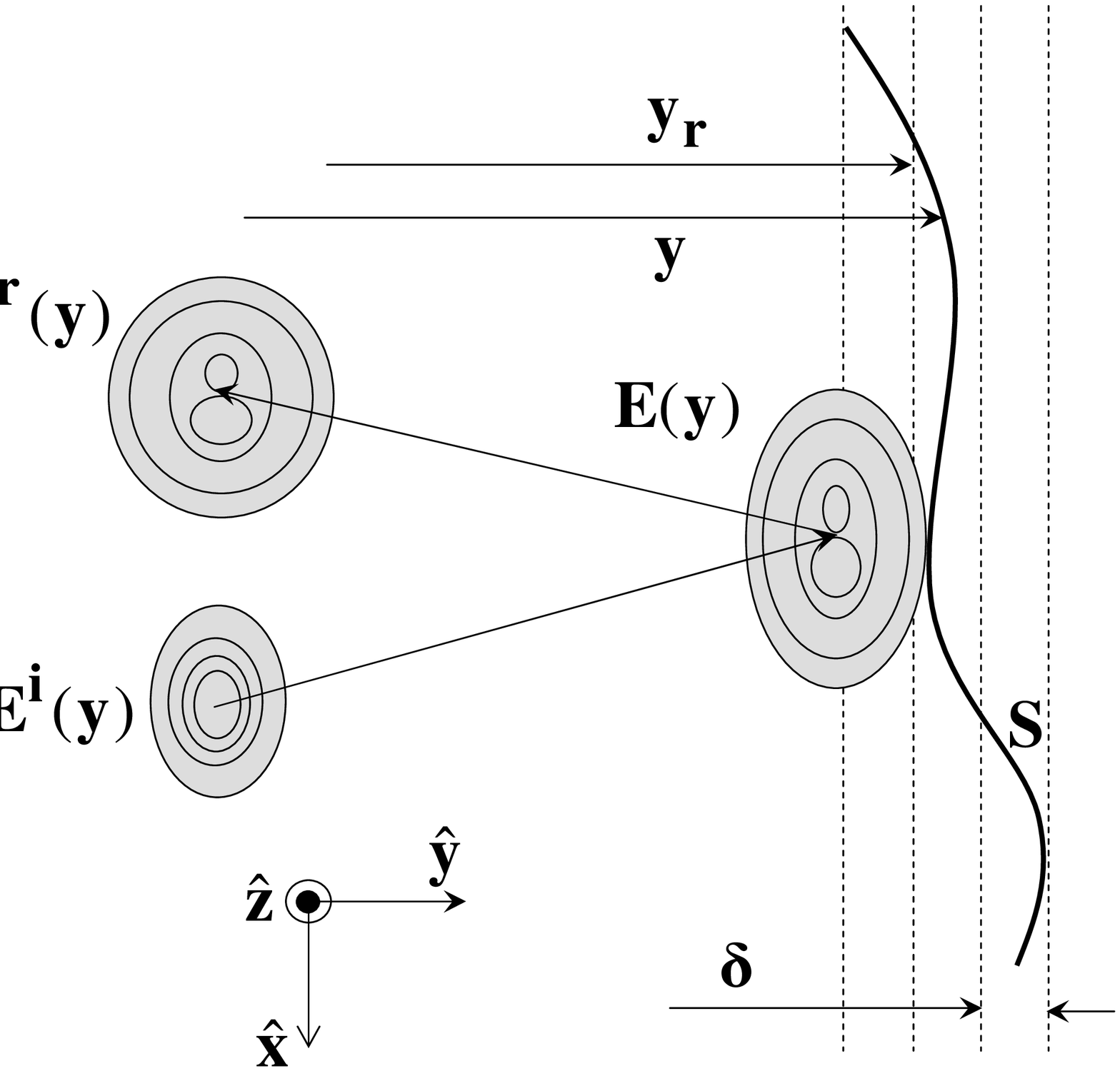}
\caption{Illustration of the   FFT-based efficient algorithm for the
phase gradient method.  ${\bf E}^i$ and  ${\bf E}^r$ are the
incident electric field and reflected electric field respectively.
$S$ is the PEC mirror phase corrector. $y$ is the coordinate of the
PEC mirror phase corrector and $y_r$ is the coordinate of the
slicing reference plane. $\delta$
 is the spacing between two adjacent slicing reference planes.
}\label{cTIFFT}
\end{figure}

  \begin{center}
  {\bf V. An  Efficient Algorithm for Phase Gradient Method  }
\end{center}

By slicing the PEC mirror phase corrector into many subdomains, as
shown in Fig. \ref{cTIFFT}, the FFT can be used
\cite{Shaolin_30,Shaolin_JEMWA} to compute   the electric field
${\bf E}$ and it's derivatives,

\vspace{-0.1in}

  \begin{eqnarray}\label{EF}
  {\bf E} (y) = \hbox{IFT} \left\{   \hspace{-0.12in} \begin{array}{cc} \\  \\ \end{array}
  {\bf F} (k_x, k_z) e^{-i k_y y}    \hspace{-0.12in} \begin{array}{cc} \\  \\ \end{array}
   \right\}, \ \ \   {\bf F} (k_x, k_z) = \hbox{FT} \left\{  \hspace{-0.12in} \begin{array}{cc} \\  \\ \end{array}
  {\bf E} (y=0)     \hspace{-0.12in} \begin{array}{cc} \\  \\ \end{array}
   \right\}
\end{eqnarray}

  \begin{eqnarray}\label{der}
  \frac{\partial {\bf E} (y)}{\partial v} =   \hbox{IFT} \left\{   \hspace{-0.12in} \begin{array}{cc} \\  \\ \end{array}
  -i k_{v} {\bf F} (k_x, k_z) e^{-i k_y y}    \hspace{-0.12in} \begin{array}{cc} \\  \\ \end{array}
   \right\}, \ \ \ \ \ v= x,z
\end{eqnarray}

\hspace{-0.2in}where, the Fourier Transform (FT) and the Inverse
Fourier Transform (IFT) are defined as follows,

  \begin{eqnarray}\label{FT}
   \hbox{FT} \left\{   \hspace{-0.12in} \begin{array}{cc} \\   \end{array}
  \cdot   \hspace{-0.12in} \begin{array}{cc} \\    \end{array}
   \right\}  & \equiv & \frac{1}{2 \pi} \int_{-\infty}^\infty dx  e^{i k_x x}
   \int_{-\infty}^\infty  \left\{   \hspace{-0.12in} \begin{array}{cc} \\   \end{array}
  \cdot   \hspace{-0.12in} \begin{array}{cc} \\    \end{array}
   \right\}      e^{i k_z z}   dz
   \end{eqnarray}

\vspace{-0.1in}

  \begin{eqnarray}\label{IFT}
 \hbox{IFT} \left\{   \hspace{-0.12in} \begin{array}{cc} \\   \end{array}
  \cdot   \hspace{-0.12in} \begin{array}{cc} \\    \end{array}
   \right\}  & \equiv & \frac{1}{2 \pi} \int_{-\infty}^\infty dk_x  e^{-i k_x x}
   \int_{-\infty}^\infty  \left\{   \hspace{-0.12in} \begin{array}{cc} \\   \end{array}
  \cdot   \hspace{-0.12in} \begin{array}{cc} \\    \end{array}
   \right\}      e^{-i k_z z}   dk_z
\end{eqnarray}

The wrapped phase difference  $\delta \check{\theta} =  \left(
\check{\theta}^i - \check{\theta}^r \right)$   is obtained from
(\ref{EF}). Due to similarity, only x-component $\hat{\bf x}
\hbox{E}_x$ is considered here,

  \begin{eqnarray}\label{derwrap}
\delta \check{\theta}_x =  \arctan\left[ \frac{ \Im \left(
\hspace{-0.12in}
\begin{array}{cc} \\  \\  \end{array} \hbox{IFT} \left\{ \hbox{
F}^r_x (k_x,k_z) e^{-i k_y y}
\right\} \hspace{-0.12in} \begin{array}{cc}  \\  \\
\end{array} \right)}{\Re\left(
\hspace{-0.12in}
\begin{array}{cc} \\  \\  \end{array} \hbox{IFT} \left\{ \hbox{
F}^r_x (k_x,k_z) e^{-i k_y y}
\right\} \hspace{-0.12in} \begin{array}{cc}  \\  \\
\end{array} \right)} \right] -   \arctan \left[ \frac{ \Im \left(
\hspace{-0.12in}
\begin{array}{cc} \\  \\  \end{array} \hbox{IFT} \left\{ \hbox{
F}^i_x (k_x,k_z) e^{-i k_y y}
\right\} \hspace{-0.12in} \begin{array}{cc}  \\  \\
\end{array} \right)}{\Re\left(
\hspace{-0.12in}
\begin{array}{cc} \\  \\  \end{array} \hbox{IFT} \left\{ \hbox{
F}^i_x (k_x,k_z) e^{-i k_y y}
\right\} \hspace{-0.12in} \begin{array}{cc}  \\  \\
\end{array} \right)} \right]
\end{eqnarray}

With the help of (\ref{EF})-(\ref{IFT}), the  gradient of the phase
difference $\nabla \left(\delta \theta_x \right)$ on the slicing
reference plane $y_r$ in Fig. \ref{cTIFFT} can be obtained from
(\ref{delta_phase}),

\begin{eqnarray}\label{diff}
\nabla \left(\delta \theta_x \right) =  \nabla \left(  \delta
\check{\theta}_x \right) = \Re \left[  \frac{ \hbox{IFT} \left\{
{\bf k} \hbox{ F}^i_x (k_x,k_z) e^{-i k_y y} \right\} }{
  \hbox{IFT} \left\{ \hbox{
F}^i_x (k_x,k_z) e^{-i k_y y} \right\}  } -   \frac{ \hbox{IFT}
\left\{ {\bf k} \hbox{ F}^r_x (k_x,k_z)  e^{-i k_y y} \right\} }{
  \hbox{IFT} \left\{ \hbox{
F}^r_x (k_x,k_z) e^{-i k_y y} \right\}  }  \right]
\end{eqnarray}

To obtain the PEC mirror surface correction $\delta_{\hat{\bf y}}$
through (\ref{phase_gradient}),  $\delta \check{\theta}_x$ has to be
unwrapped. Here, an FFT-based phase unwrapping algorithm is
presented for the $r$-norm minimum problem given in (\ref{norm}).
Suppose that $\delta \theta_x$ can be expressed in the Fourier
series,

\begin{eqnarray}\label{Fourier}
\delta \theta_x =  \hbox{IFT} \{ f (k_x,k_z)  \}
\end{eqnarray}

Then,

\begin{eqnarray}\label{Fourierder}
 \frac{\partial \left( \delta \theta_x \right) }{\partial v} =  \hbox{IFT} \{ -i k_v  f (k_x,k_z)
 \},  \ \ \ \ \ v=x, z
\end{eqnarray}

To obtain $\delta \theta_x$,   the Fourier coefficient $f (k_x,k_z)$
is chosen to minimize the cost function $Q$ given in (\ref{norm}),
with $w_x = w_z =1$.  For LMS method where $r = 2$, it can be shown
that $f (k_x,k_z)$ takes the following form,

\begin{eqnarray}\label{f}
 \frac{ Q \left(f + \delta f \right) -  Q \left(f  \right)}{\delta f} = 0 \rightarrow  f (k_x,k_z) = i  \frac{k_x
  \hbox{FT} \left\{  \nabla \left( \delta \check{\theta}_x  \right) \cdot \hat{\bf x}    \right\}
  +  k_z  \hbox{FT}\left\{ \nabla \left( \delta \check{\theta}_x \right) \cdot \hat{\bf z}   \right\} }{k_x^2 + k_z^2 }
\end{eqnarray}

Now, the PEC mirror surface correction $\delta_{\hat{\bf y}}$ can be
obtained from (\ref{phase_gradient}), with the help of
(\ref{diff})-(\ref{f}).

  \begin{center}
  {\bf VI. Discussion }
\end{center}

It has been shown that both the GO method and the phase gradient
method can be used in the PEC mirror phase corrector design. Both
methods have their advantages and disadvantages, e.g., the GO method
is simple and easy to use, but it is time-consuming; the phase
gradient method is efficient (due to the use of FFT), but it's
application is limited by the sampling theorem. In general, the GO
method is suitable for problems of significant side lobes; and the
phase gradient method is suitable for problems of smooth phase front
with negligible side lobes.

 \begin{center}
  {\bf VII. Conclusion }
\end{center}

In this article, both the GO method and the phase gradient method
have been presented for the PEC mirror phase corrector design. The
FFT-based efficient algorithm has been proposed for the phase
gradient method to speed up the design procedure.

  %\medskip

  \begin{center}
  {\bf Acknowledgements }
\end{center}

This work was supported by the U.S. Dept. of Energy under the
contract DE-FG02-85ER52122.

\begin{center}

\end{center}

\end{paper}
%--------------------------

\end{document}